\documentclass[conference]{IEEEtran}
\IEEEoverridecommandlockouts
\usepackage{cite}
\usepackage{url}
\usepackage{xcolor}
\usepackage{comment}
\usepackage{array}
\usepackage{amsmath,amssymb,amsfonts}
\usepackage{graphicx}
\usepackage{textcomp}
\usepackage{xcolor}
\usepackage{algorithm}
\usepackage{algpseudocode}
\usepackage{mathptmx}
\usepackage{multirow}
\usepackage{booktabs}
\usepackage{color,soul}
\usepackage{amsfonts}
\def\BibTeX{{\rm B\kern-.05em{\sc i\kern-.025em b}\kern-.08em
    T\kern-.1667em\lower.7ex\hbox{E}\kern-.125emX}}
\begin{document}
\title{Uncertainty Error Modeling for Non-Linear State Estimation With Unsynchronized SCADA and $\mu$PMU Measurements\\

}
\author{
\IEEEauthorblockN{Austin Cooper}
\IEEEauthorblockA{\textit{Dept. of ECE} \\
\textit{University of Florida}\\
Gainesville, FL, USA.\\austin.cooper@ufl.edu}
\and

\IEEEauthorblockN{Arturo Bretas}
\IEEEauthorblockA{\textit{Distributed Systems Group}\\
\textit{Pacific Northwest National Laboratory}\\
Richland, WA, USA.\\arturo.bretas@pnnl.gov}
\and

\IEEEauthorblockN{Sean Meyn}
\IEEEauthorblockA{\textit{Dept. of ECE} \\
\textit{University of Florida}\\
Gainesville, FL, USA.\\meyn@ece.ufl.edu}
\and

\IEEEauthorblockN{Newton G. Bretas}
\IEEEauthorblockA{\textit{Dept. of ECE} \\
\textit{University of São Paulo}\\
São Carlos, SP, BRA.\\ngbretas@sc.usp.br}
}

\maketitle

\begin{abstract}
Distribution systems of the future smart grid require enhancements to the reliability of distribution system state estimation (DSSE) in the face of low measurement redundancy, unsynchronized measurements, and dynamic load profiles. Micro phasor measurement units ($\mu$PMUs) facilitate co-synchronized measurements with high granularity, albeit at an often prohibitively expensive installation cost. Supervisory control and data acquisition (SCADA) measurements can supplement $\mu$PMU data, although they are received at a slower sampling rate. Further complicating matters is the uncertainty associated with load dynamics and unsynchronized measurements—not only are the SCADA and $\mu$PMU measurements not synchronized with each other, but the SCADA measurements themselves are received at different time intervals with respect to one another. This paper proposes a non-linear state estimation framework which models dynamic load uncertainty error by updating the variances of the unsynchronized measurements, leading to a time-varying system of weights in the weighted least squares state estimator. Case studies are performed on the 33-Bus Distribution System in MATPOWER, using Ornstein–Uhlenbeck stochastic processes to simulate dynamic load conditions.
\end{abstract}


\section{Introduction}\label{sec:intro}
Increasing power distribution system (PDS) complexity and distributed energy resource (DER) penetration present complicated challenges for the modern smart grid. As a consequence, great effort has been dedicated to bolster the reliability of distribution management systems (DMS) given the growing uncertainty associated with increasingly complicated load dynamics and the much improved, but nevertheless imperfect, synchronization of metering equipment \cite{distsurvey}. 

This uncertainty error associated with unsynchronized measurements most significantly affects the distribution level, where non-observability issues due to measurement device scarcity at the low and medium voltage levels \cite{dssehistory} further complicate obtaining a reliable snapshot of the system. This directly hinders distribution system state estimation (DSSE), which relies on system observability to ensure state variable estimation from the measurement set \cite{monticelli}. Pseudo measurements have been proposed to increase system observability artificially \cite{pseudo}, however poor accuracy and reliability of these measurements may corrupt the DSSE, especially with the increasing complexity of PDS and DER penetration. 

To compensate, advanced metering infrastructure (AMI) devices have been proposed to augment the measurement set. The work in \cite{AMI} proposes a DSSE using AMI, although the measurements are considered synchronized. Supervisor control and data acquisition (SCADA) and micro-phasor measurement units ($\mu$PMUs) have been used with AMI in \cite{Rodrigo} which, like the proposed work, models time-varying loads as Ornstein–Uhlenbeck stochastic processes. The referenced work uses linear state estimation (LSE) based on complex current and voltage measurements, leveraging the high granularity of $\mu$PMUs to quickly obtain state variable estimates \cite{pmubook}. State estimation (SE) aims to obtain the complex voltages at each system bus based on measurement data, which is instrumental to power system monitoring \cite{bretas2021cyber}. In the referenced LSE approach, measurement variances are then updated based on the OU load model \cite{whitenoise}, improving SE results. Shortcomings of this implementation include high noise sensitivity. Further, the LSE's necessary linearization of non-linear SCADA and AMI measurement functions introduces additional potential for error. The proposed model seeks to address these issues.

The idea for this work is to incorporate the uncertainty error associated with unsynchronized measurements into a two-step non-linear SE measurement model, rather than deciphering the cause of uncertainty error after the fact. The few high-granularity $\mu$PMU measurements will be used to obtain synchronized active and reactive power injections at the $\mu$PMU buses. Unsynchronized SCADA measurements, including active and reactive power flows and injections, will not require linearization. Case studies will explore various redundancy levels and SCADA sampling rates. Further, the co-asynchronicity among SCADA will be simulated by staggering their arrival times with respect to one another.

Three specific contributions of this paper towards the state-of-the-art include:
\begin{itemize}
    \item An enhanced two-step non-linear SE framework which includes uncertainty error in the measurement model. 
    \item A formal modeling of the LSE truncation error, which the proposed work circumvents.
    \item Robustness to the asynchronicity between SCADA and $\mu$PMU measurements, as well as the co-asynchronicity between SCADA. This is in contrast to traditional methods which treat measurements as synchronized.
\end{itemize}

   \begin{figure}%[h]
\includegraphics[width=7cm,height = 8.708cm,scale = 1]{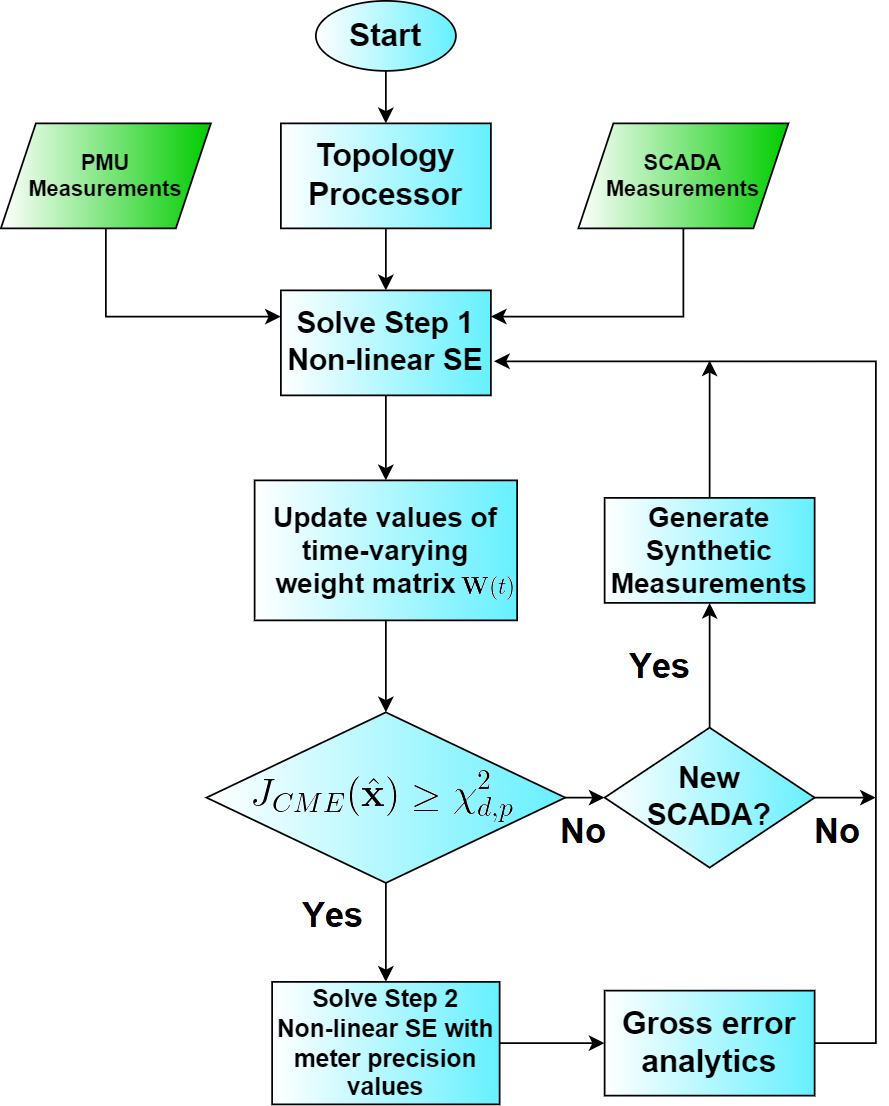}
\centering
\caption{ Proposed Uncertainty Modeling Non-linear SE Flowchart.
}
\label{flowchart}
\end{figure}

\section{Background}\label{sec:background}
\subsection{Non-Linear State Estimation}\label{sec:SEbackground}
Classical power system SE follows a Weighted Least Squares (WLS) framework \cite{bretas2021cyber}. A power system with $n$ buses and $d$ measurements can be modeled as a set of non-linear algebraic equations in the following measurement model:
\begin{equation}
    \mathbf{z} = h(\mathbf{x}) + \mathbf{e}
\label{eq:SE}\end{equation}
where $\mathbf{z}\in\mathbb{R}^{1 \times d}$ is the measurement vector, $\mathbf{x}\in\mathbb{R}^{1 \times N}$ is the state variables vector, $h:\mathbb{R}^{1 \times N}\rightarrow\mathbb{R}^{1 \times d}$ is a continuous non-linear differentiable function, and $\mathbf{e}\in\mathbb{R}^{1 \times d}$ is the measurement error vector.  Each measurement error $e_i$ is assumed to follow a zero mean Gaussian distribution.  $d$ is the number of measurements while $N=2n-1$ is the number of unknown state variables, i.e., the complex voltages at each bus.

In the traditional WLS approach, the best state vector estimate in \eqref{eq:SE} is determined by minimizing the weighted norm of the residual, represented with the cost function $J(\mathbf{x})$:
\begin{equation} 
\label{eq:JSE}
J(\mathbf{x})=\Vert \mathbf{z}-h(\mathbf{x})\Vert _{\mathbf{W}}^{2}=[\mathbf{z}-h(\mathbf{x})]^{T}\mathbf{W}[\mathbf{z}-h(\mathbf{x})] 
\end{equation}
where $\mathbf{W}$ is the inverse covariance matrix of the measurements, otherwise known as the weight matrix. The solution to \eqref{eq:JSE} is found through the iterative Newton-Raphson method. The linearized form of \eqref{eq:SE} then becomes:

\begin{equation} \Delta \mathbf{z}=\mathbf{H}\Delta \mathbf{x}+\mathbf{e} \label{SElin} \end{equation}
where $\mathbf{H}=\frac{\partial h}{\partial \mathbf{x}}$ is the Jacobian matrix of $h$ at the current state estimate $\mathbf{x}^*$, $\Delta \mathbf{z}=\mathbf{z}-h(\mathbf{x}^*)=\mathbf{z}-\mathbf{z}^*$ is the measurement vector correction and $\Delta \mathbf{x}=\mathbf{x}-\mathbf{x}^*$ is the state vector correction.  The WLS solution can further be understood geometrically \cite{geom} as the projection of $\Delta \mathbf{z}$ onto the Jacobian space $\mathfrak{R}$($H$) by a linear projection matrix $\mathbf{K}$, i.e. $\Delta\hat{\mathbf{z}}=\mathbf{K}\Delta\mathbf{z}$:
\begin{equation}\label{eq:P}
\mathbf{K} = \mathbf{H}(\mathbf{H}^{T}\mathbf{W}\mathbf{H})^{-1}\mathbf{H}^{T}\mathbf{W}.
\end{equation}

The SE can further be interpreted by visualizing the geometrical position of the measurement error in relation to the Jacobian range space $\mathfrak{R}$($H$). Through decomposing the measurement vector space into a direct sum of $\mathfrak{R}$($H$) and $\mathfrak{R}($H$)^{\perp}$, it is possible to also decompose the measurement error vector $\mathbf{e}$ into two components:
\begin{equation} \label{eq:e}
\mathbf{e} = \underbrace{\mathbf{K}\mathbf{e}}_{\mathbf{e_U}} + \underbrace{(I-\mathbf{K})\mathbf{e}}_{\mathbf{e_D}}.
\end{equation}

The component $\mathbf{e_D}$ is the detectable error, which is equivalent to the residual in the classical WLS model. The component $\mathbf{e_U}$ is the undetectable component of the residual. $\mathbf{e_D}$ lies in the space orthogonal to the Jacobian range space, while $\mathbf{e_U}$ is hidden in the Jacobian space \cite{bretas2021cyber}. To formalize the impact of the undetectable component $\mathbf{e_U}$, the measurement Innovation Index ($II$) is introduced \cite{II}:
\begin{equation}\label{eq:10}
{II}_{i} = \frac{\Vert{e^i_D}\Vert_{\mathbf{R}^{-1}}}{\Vert{e^i_U}\Vert_{\mathbf{R}^{-1}}} = \frac{\sqrt{1-K_{ii}}}{\sqrt{K_{ii}}}.
\end{equation}

A low $II$ indicates that a large component of the error is not reflected in the residual. The composed measurement error ($CME$) can then be formulated in terms of the residual and $II$, after which the normalized $CME^N$ is obtained:
\begin{equation}\label{eq:11}
CME_i = r_i\left(\sqrt{1+\frac{1}{{II_i}^2}}\right) \Rightarrow CME_i^N = \frac{CME_i}{\sigma_i}
\end{equation}
where $\sigma_i$ is the $i$-th measurement standard deviation.

The SE yields $CME$ values based on measurements taken throughout the power system. Real measurements in this work are defined as those obtained from SCADA and $\mu$PMUs. In addition,  per the methodology in \cite{synth}, synthetic measurements (SM) are artificially created in low redundancy areas considering measurement $II$ and $n$-tuple of critical measurements. This serves to maintain a high global redundancy level ($GRL$), i.e., the number of measurements divided by the number of state variables, by increasing the $\chi^2$ degrees of freedom, improving the reliability of the $\chi^2$ hypothesis test \cite{bretas2021cyber}.

$\chi^2$ hypothesis testing is used for bad data detection during gross error (GE) analytics. The $CME$-based objective function value (\ref{eq:chi-squared}) is compared to a $\chi^2$ threshold. The value of this threshold depends on a chosen probability $p$ and the degrees of freedom $d$, taken to be the number of measurements fed to the SE:
\begin{equation} \label{eq:chi-squared}
J_{CME}(\hat{\mathbf{x}})=\sum_{i=1}^d \left[\frac{CME_i}{\sigma_i}\right]^2 \ge \chi^2_{d,p}.
\end{equation}

If the value of $J_{CME}$ is greater than or equal to the $\chi^2$ threshold, then a GE is detected. For this work, a two-step SE approach is used, illustrated in Fig. \ref{flowchart}. The classical first step presented in \cite{twostep} uses an empirical approach wherein all measurements are weighted equally proportional to the measurement magnitude, after which the $J_{CME}(\hat{\mathbf{x}})$ is compared to $\chi^2_{d,p}$. Only if $J_{CME}(\hat{\mathbf{x}}) \geq \chi^2_{d,p}$ does the procedure advance to the second SE step, during which the SE is repeated—this time with meter precision values for each measurement type per state-of-the-art methodologies \cite{monticelli}. This serves to strike a balance between GE detectability and deciphering the measurement(s) in error. Next, following \cite{ParameterProof}, bad data is identified through analysis of the $CME^N$. This work seeks to improve upon the first-step of the classical two-step SE by replacing the empirical assumption with a continually updating model of the unsynchronized measurement uncertainty error.

\subsection{Dynamic Load Modeling}\label{sec:dynloadmodel}

Due to the asynchronicity between $\mu$PMU and SCADA measurements, as well as the co-asynchronicity among SCADA, load dynamics cause SCADA measurements to become obsolete, thus providing an inaccurate snapshot of the system state. This uncertainty error must be included into the error vector $\mathbf{e}$ from (\ref{eq:SE}) to better provide accurate state variables, error detection, and GE analytics. The drift in the system state due to load dynamics can be modelled as Ornstein–Uhlenbeck (OU) stochastic processes, which have been used to successfully reflect time-varying load profiles \cite{OUvalidate}. OU load modelling relies on the assumption that, although load demand profiles do fluctuate over time, they rarely deviate sharply from their previous value. Further, the OU is a mean-reverting process, making it an appropriate model for ascribing the uncertainty error to outdated measurement variance only.

\begin{figure}%[h]
\centering
\includegraphics[width=8.9cm,height = 4.47789cm,scale = 1]{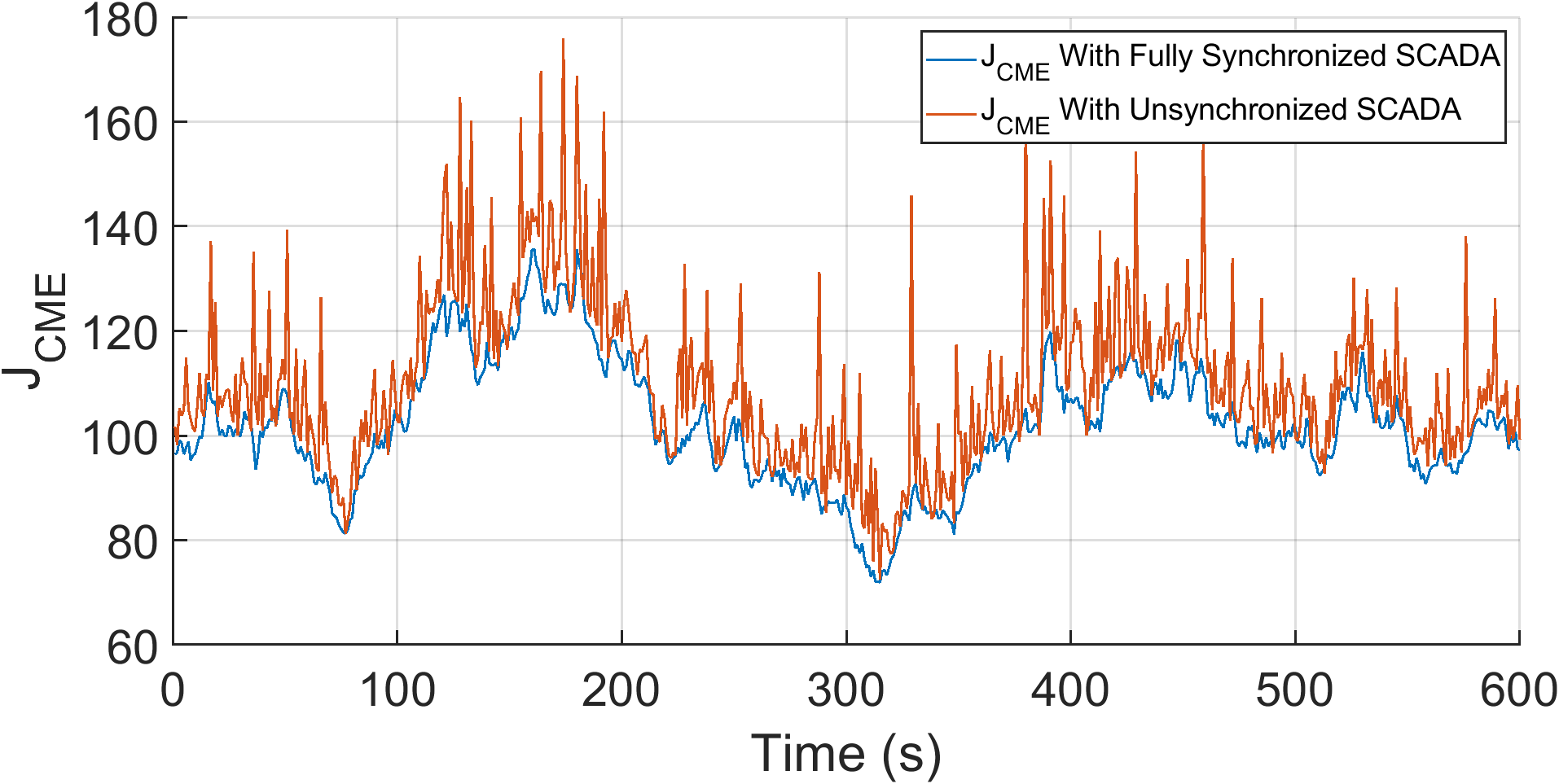}
\caption{Comparison of $J_{CME}$ paths when SCADA measurements are synchronized (blue) and unsnychronized with each other (orange).}
\label{Jc_synch_unsynch}
\end{figure}

\begin{figure}%[h]
\centering
\includegraphics[width=8.9cm,height = 4.47789cm,scale = 1]{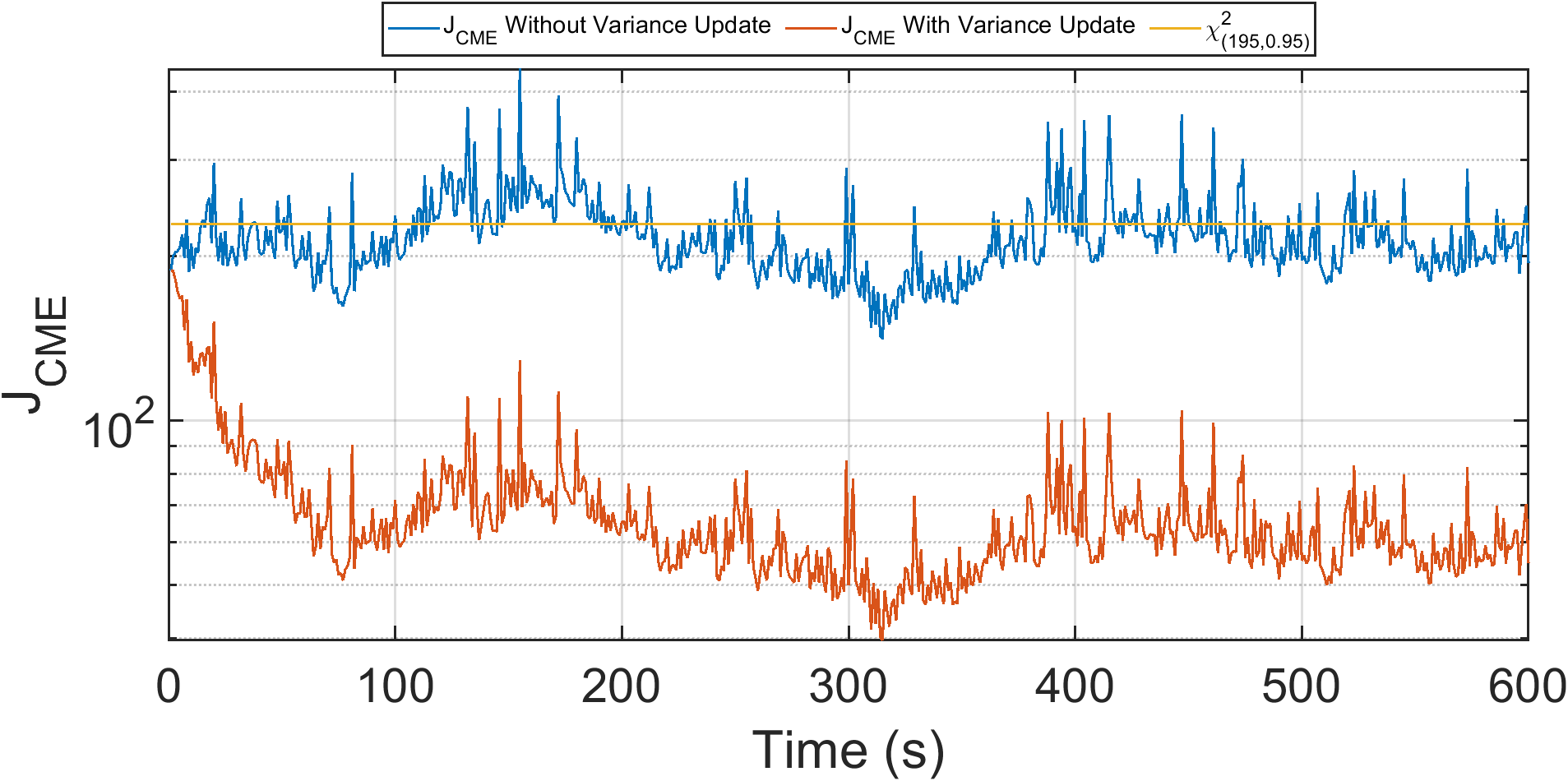}
\caption{Comparison of $J_{CME}$ paths when weight matrix $\mathbf{W}$ values are static (blue) and when $\mathbf{W}(t)$ values are updated (orange).}
\label{Jc}
\end{figure}

A stochastic differential equation is used to represent the $i$-th load apparent power, $S_i(t) = P_i(t) + jQ_i(t)$, where $P_i(t)$ and $Q_i(t)$ are the $i$-th load active and reactive power respectively:

\begin{equation} \label{eq:diffeq}
dS_i(t) = -\theta_i S_i(t)dt + \sigma_iW_i(t)
\end{equation}
where $\theta_i$ $>$ 0 is the decay rate controlling to what degree the load varies from its initial value, $\sigma_i$ $>$ 0 controls the scale of the noise, and $W_i(t)$ denotes a complex Wiener process whose sample paths follow a continuous Gaussian process \cite{wind}. Statistical analysis of $\mu$PMU data can be used to estimate these parameters \cite{OUvalidate}. The load variations can then be expressed as a discrete time process for every $\Delta$t $\mu$PMU sample \cite{Rodrigo}:

\begin{equation} \label{eq:discreteform}
S_i[k] = S_i(\tau)e^{-\theta_i\Delta t(k-q)}+\zeta_i
\end{equation}
where $\tau$ $<$ t is the time when unsynchronized SCADA is received, $t=k\Delta t$ and $\tau = q\Delta t$ for $k$ and $q$ time indices, and $\zeta_i$ is a complex random variable of zero mean and variance $\frac{\sigma_i^2}{2\theta_i}(1-e^{-2\theta_i(t-\tau)})$. Dynamic load uncertainty can substantially increase the $J_{CME}(\hat{\mathbf{x}})$ in the absence of actual measurement error, especially when measurements are unsynchronized, as illustrated in Fig. \ref{Jc_synch_unsynch}. The variance of the discrete-time Gaussian variable $S_i[k]$ can then be found recursively by:

\begin{equation} \label{eq:updateEq}
\mathrm{Var}\bigr [S_i[k]\bigr ] = \mathrm{Var}\bigr [S_i[k-1]\bigr ]\gamma + \frac{\sigma_i^2}{2\theta_i}(1-\gamma )
\end{equation}
where $\gamma = e^{-2\theta_i \Delta t}$. The updated variances of the unsynchronized SCADA measurements then constitute the time-varying inverse covariance matrix of the measurements, $\mathbf{W}(t)$. The improvement in the $J_{CME}(\hat{\mathbf{x}})$ due to the updating $\mathbf{W}(t)$ is observed in Fig. \ref{Jc}, as the $J_{CME}(\hat{\mathbf{x}})$ should not exceed the $\chi^2$ threshold due to load dynamics only.

\subsection{Non-Linear Measurement Uncertainty}\label{nonlinuncert}

The classical WLS model in (\ref{eq:SE}) does not consider the possibility of errors attributed to dynamic load measurement uncertainty. This model can be augmented by considering, for any $i$-th measurement, a noiseless true measurement $z_i^{true}[k]$ at time $k$, which is related to a previous value $z_i^{true}[q]$ plus random load variation $\Delta z_i^{true}[k]$, given by (\ref{eq:trueEq}), whereas the noisy and out-of-date unsynchronized measurement, $z_i^{meas}[q]$, is represented in (\ref{eq:timemeasEq}):

\begin{equation} \label{eq:trueEq}
z_i^{true}[k] = z_i^{true}[q] + \Delta z_i^{true}[k]
\end{equation}
\begin{equation} \label{eq:timemeasEq}
 z_i^{meas}[q] = z_i^{true}[q] + e_i[q]
\end{equation}
where $e_i[q]$ is the measurement noise of the same Gaussian distribution as in (\ref{eq:SE}). At a given instance $k$ when the first step of the non-linear SE is performed, the precise value of $z_i^{true}[k]$ is unobtainable; the best available measurement in time is $z_i^{meas}[q]$. However, it is possible to approximate $z_i^{meas}[k]$ by assuming that $\Delta z_i^{true}[k]$ is attributed \emph{only} to load variations:
\begin{equation} \label{eq:third}
z_i^{meas}[k] = z_i^{meas}[q] + \Delta z_i^{true}[k]
\end{equation}

The model in (\ref{eq:third}) is used for approximating the unsynchronized SCADA measurement variances, which are used to update the time-varying $\mathbf{W}(t)$ weight matrix.

The proposed work also seeks to formalize its assertion of preserving measurement fidelity. The work in \cite{Rodrigo} uses an OU process load modeling approach with variance update equations, however it relies on linearizing SCADA and AMI measurements. Unlike the proposed work, this linearization process introduces yet an additional error component exclusive to the LSE approach, which this work calls the linearization truncation error $\epsilon^l$. To formalize this additive error, the Taylor series of the LSE measurement model can be developed as:
\begin{equation} \label{eq:trunc}
z_i = h_{i,0} + \frac{\partial h_i}{\partial x}\Delta x - \epsilon_i^l
\end{equation}

By setting (\ref{eq:trunc}) equal to the standard SE measurement model $z_i = h_i(x) + e_i$, one obtains:
\begin{equation} \label{eq:trunc2}
h_i(x) + e_i = h_{i,0} + \frac{\partial h_i}{\partial x}\Delta x - \epsilon_i^l
\end{equation}

By defining the linearized form of $z_i$ as $z_i^l = h_{i,0} + \frac{\partial h_i}{\partial x}\Delta x$, one finally obtains:
\begin{equation} \label{eq:trunc3}
z_i^l = h_i(x) + e_i + \epsilon_i^l
\end{equation}
where $e_i$ is the measurement error and $\epsilon_i^l$ is the LSE additive truncation error, which contributes to SE inaccuracies. The proposed model does not require measurement function linearization, and thus preserves measurement fidelity by avoiding the error component $\epsilon_i^l$. Further, in the referenced LSE approach, additional error is introduced in the calculation of linearized measurement covariances, which must be approximated using the truncated Taylor series of their functions \cite{Rodrigo}. The proposed work thus further preserves measurement fidelity by circumventing measurement covariance approximation.

\section{Case Study}\label{casestudy}
Validation was performed in MATPOWER on the 12.66 kV 33-bus distribution system from Baran and Wu \cite{33bus}. OU processes were used to model each load using a $\theta$ value of 0.0125, as obtained from \cite{OUvalidate}. An optimal number of 11 $\mu$PMUs was selected per the methodology in \cite{PMUoptimal}. SCADA sampling rates of 1, 2, and 4 seconds were considered. To simulate asynchronicity, the SCADA measurements were organized into 33 groups. If a $\mu$PMU is present at a given bus, then only the active and reactive power flows to and from that bus are considered as SCADA measurements for that group. Otherwise, the active and reactive power injections are also considered as SCADA for that group. The remaining power injections are obtained through the $\mu$PMUs or SM. 

For a given SCADA sampling rate, maximum asynchronicity was simulated by staggering the SCADA arrival rates as far apart as possible by a factor of integer division $\lfloor f_{PMU}/N_{groups} \rfloor$, where $f_{PMU}$ is the $\mu$PMU sampling rate and $N_{groups}$ are the number of SCADA groups. To test performance with smaller data sets, two separate measurement plans were considered of 180 and 195 measurements ($GRL = 3$ and $GRL = 2.769$ respectively). $p$ = 0.95 was considered for the $\chi^2$ threshold, however this can be customized depending on desired dependability/security trade-off. Random errors with distribution $X \sim \mathcal{N}(0,1)$ were applied to all measurements.

The first SE iteration is performed once SCADA information from every group is collected, along with the $\mu$PMU data. In the classical two-step SE approach described in Section \ref{sec:SEbackground}, an empirical approach is considered in which all measurements are considered as possibly having GEs, with each measurement standard deviation being a percentage of the measurement magnitude $(\sigma_i = |z_i|/100)$ \cite{convprop}. In the proposed work, however, the classical empirical approach is now replaced with the uncertainty error modeling embedded into the time-varying weight matrix $\mathbf{W}(t)$, which is continually updated using the variances obtained from (\ref{eq:updateEq}).

If left unmodeled, complications of the added uncertainty error to the error vector $\mathbf{e}$ include: 1.) erroneous triggering of protection and control devices \cite{relay} due to the $J_{CME}$ value exceeding the $\chi^2$ threshold—a process that should be reserved for detecting measurement errors \cite{bretas2021cyber} during GE analytics—and 2.) reduced state estimate accuracy. Therefore, the case study will focus on these two metrics.

To quantify improved state variable accuracy, the metric $SE_{error}$ is defined as the L2-norm of the estimate error:
\begin{equation} 
\label{eq:error}
SE_{error} = \sqrt{\sum_{i=1}^{B}(v_i^{act} - \hat v_i)^2}
\end{equation}
where $B$ is the number of buses, $v_i^{act}$ is the actual voltage magnitude at the $i$-th bus with zero error, and $\hat v_i$ is the estimated voltage magnitude at the $i$-th bus.

Three cases were explored for this analysis. An ``Ideal SE" case considers perfect synchronicity across $\mu$PMU and SCADA, meaning that the only error introduced into the SE process comes from measurement noise; uncertainty error is absent. Although unrealistic, the ideal case provides comparative insight between the traditional and proposed unsynchronized models. The ``Traditional SE" case considers asynchronicity across all measurements with a static weight matrix $\mathbf{W}$. Finally, the ``Proposed SE" employs the updating time-varying weight matrix $\mathbf{W}(t)$. Each simulation was run for 6 hours in MATPOWER with varying OU load dynamics.

\begin{figure}%[h]
\centering
\includegraphics[width=7.05cm,height = 5.9029385625cm,scale = 1]{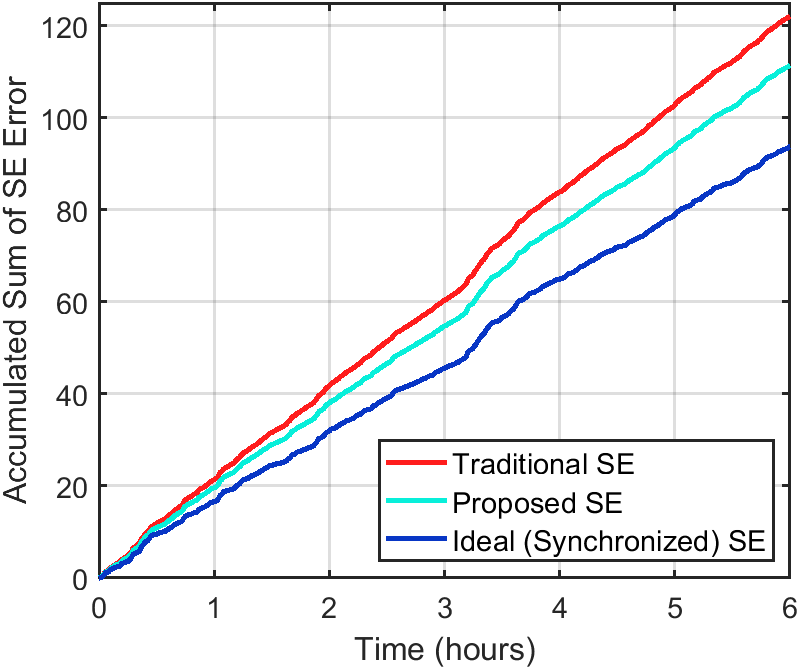}
\caption{Comparison of state variable estimate accuracy}
\label{errorPlot}
\end{figure}

Fig. \ref{errorPlot} illustrates the improvement of the proposed model's state variable accuracy by plotting the cumulative sum of the $SE_{error}$ metric, showing a 37.7746\% decrease in the $SE_{error}$ for moderate SCADA asynchronicity. As expected, the Proposed SE improved upon state variable accuracy by updating the measurement variances, as opposed to the Traditional SE which contains no such uncertainty modeling.

Next, the improvement in false positive rate (FPR) of the proposed model was evaluated. FPR refers to the erroneous triggering of the GE analytics process, which could result in a measurement being falsely deleted from the measurement set per standard SE practices \cite{bretas2021cyber}, hindering the reliability of $\chi^2$ hypothesis testing. In a worse case scenario, this could lead to false activation of the protection and control apparatus \cite{relay}. Table I compares the FPRs across three different SCADA sampling rates at maximum asynchronicity and a $GRL$ of 3, where Table II reduces the measurement set to a $GRL$ of 2.769. A substantial decrease in FPR was observed with variance updating, which in turn reduced the proportion of uncertainty error in the error vector $\mathbf{e}$.

\begin{table}[h]
\caption{Gross Error FPR Comparison for GRL = 3}
\begin{center}
\renewcommand{\arraystretch}{1.1}
\begin{tabular}{ |c|c|c| } 
  \hline
  SCADA Sampling Rate & FPR Without Update & \textbf{FPR With Update} \\ 
  \hline
  1 s & 1.32\% & \textbf{0.00}\% \\ 
  \hline
  2 s & 11.81\% & \textbf{0.00}\%  \\ 
  \hline
  4 s & 32.24\% & \textbf{0.98}\%  \\ 
  \hline
\end{tabular}
\end{center}

\medskip
\caption{Gross Error FPR Comparison for GRL = 2.769}
\begin{center}
\renewcommand{\arraystretch}{1.1}
\begin{tabular}{ |c|c|c| } 
  \hline
  SCADA Sampling Rate & FPR Without Update & \textbf{FPR With Update} \\ 
  \hline
  1 s & 47.17\% & \textbf{0.00}\% \\ 
  \hline
  2 s & 64.49\% & \textbf{5.23}\%  \\ 
  \hline
  4 s & 81.22\% & \textbf{12.84}\%  \\ 
  \hline
\end{tabular}
\end{center}
\end{table}

\section{Summary and Conclusions}\label{sec:conclusion}

This work presents a two-step non-linear SE framework which includes load dynamics-induced uncertainty error into the measurement model. The performance of the proposed method was evaluated with worst case scenario asynchronicity, varying degrees of unsynchronized SCADA arrival time, and a reduced input measurement data set, showcasing its reliability and practicality compared to conventional SE. Two major improvements were observed in the proposed work when compared to conventional SE: increased accuracy of state variable estimates and decreased FPR associated with erroneous triggering of the bad data detection step. Further, a formalizing of the truncation error $\epsilon^l$ associated with a previous LSE approach with OU load modeling was presented, which the proposed work circumvents by preserving the non-linear measurement functions, leading to more accurate state variable estimates and robustness to approximation error.

\bibliographystyle{IEEEtran}
\bibliography{main}

\end{document}